\documentstyle[epsf]{article}
\begin{document}
\baselineskip7mm
\title{Chaos in FRW cosmology with gently sloping scalar field potentials}
\author{S.A. Pavluchenko$^{\dagger}$  and  A.V. Toporensky$^{\ddagger}$}
\date{}
\maketitle
\hspace{8mm}{\em Sternberg Astronomical Institute,
 Moscow 119899, Russia}
\begin{abstract}
The chaotic behavior in FRW cosmology with a scalar field is studied
for scalar field potentials less steep than quadratic. We describe
a transition to much stronger chaos for appropriate parameters of
such potentials. The range of parameters which
allows this transition is specified.
The influence of ordinary matter on the
chaotic properties of this model is also discussed.
\end{abstract}
$^{\dagger}$    Electronic mail: sergey@sai.msu.su \\
$^{\ddagger}$ Electronic mail: lesha@sai.msu.su\\
\section{Introduction}
In the last few years the chaotic regime in dynamics of closed FRW universe
filled with a scalar field becomes the issue of investigations.
Initially, the model with a massive scalar field (with the scalar field
potential $V(\varphi)=(m^2 \varphi^2)/2$, where $m$ is the mass of the scalar
field) was studied \cite{Page,Corn-Shel}.  Before summarizing main
 results obtained, we present the equation of motion
(for further using we will not specify the
potential $V(\varphi)$). The system has two dynamical variables - the
scale factor $a$ and the scalar field $\varphi$:
\begin{equation}
\frac{m_{P}^{2}}{16 \pi}\left(\ddot{a} + \frac{\dot{a}^{2}}{2 a}
+ \frac{1}{2 a} \right)
+\frac{a \dot{\varphi}^{2}}{8}
-\frac{a V(\varphi)}{4} = 0,
\end{equation}
\begin{equation}
\ddot{\varphi} + \frac{3 \dot{\varphi} \dot{a}}{a}
+ V'(\varphi) = 0.
\end{equation}
with the first integral
\begin{equation}
-\frac{3}{8 \pi} m_{P}^{2} (\dot{a}^{2} + 1)
+\frac{a^{2}}{2}\left(\dot{\varphi}^{2} + 2 V(\varphi)\right)  =
0.
\end{equation}
Here $m_P$ is the Planck mass.

The points of maximal expansion
and those of minimal contraction, i.e. the points, where $\dot{a} =
0$ can exist only in the region where
\begin{equation}
a^{2} \leq \frac{3} {8 \pi}  \frac{m_{P}^2}{V(\varphi)} ,
\end{equation}
Sometimes, the region defined by inequalities (4) is called
the Euclidean one.
One can easily see also that the possible points of
maximal expansion (where $\dot a=0$, $\ddot a<0$) are localized
inside the region
\begin{equation}
a^{2} \leq \frac{1}{4 \pi} \frac{m_{P}^{2}}{V(\varphi)}
\end{equation}
while the possible points of minimal contraction (where
$\dot a=0$, $\ddot a>0$) lie outside this
region (5) being at the same time inside the Euclidean region
(4).

The main idea of the further analysis \cite{our}
consists in the fact that in the closed isotropical model with a
minimally coupled scalar field satisfying the energodominance condition
all the trajectories have the
point of maximal expansion.  Then
the trajectories can be classified according to localization of their
points of maximal expansion. The area of such points is specified by
(5).  A numerical investigation shows that this area has a quasi-
periodical structure, wide zones corresponding to the falling to
singularity being intermingled with narrow those in which the points
of maximal expansion of trajectories having the so called ``bounce'' or
point of minimal contraction are placed. Then studying the substructure
of these zones from the point of view of possibility to have two
bounces one can see that this substructure reproduce on the qualitative
level the structure of the whole region of possible points of
maximal expansion.  Continuing this procedure {\it ad infinitum}
yields the fractal set of infinitely bouncing trajectories.

It should be noticed that even the 1-st order bounce intervals (containing
maximum expansion points for trajectories having at least one bounce)
are very narrow. Analytical approximation for large initial $a$ indicates
that the width of intervals is roughly inversely proportional to $a$
\cite{Star}.
The opposite case of small initial $a$ was investigated numerically,
and the ratio of the first such interval width to the distance between
intervals appear to be of the order of $10^{-2}$, if we do not take
into account zigzag-like trajectories.  So, the chaotic regime, though
 being interesting from the mathematical point of view, may be treated
as not important enough.

For steeper potentials the chaos is even less significant. The chaotic
behavior may disappear completely for exponentially steep potentials
 \cite{our2} .

The goal of the present paper is to describe the opposite case - the potentials
which is less steep than the quadratic one. We will see that in this case the
transition to a qualitatively stronger chaos may occur.

The structure of the paper is as follows. It Sec.2 we consider
asymptotically flat potential and explain new features of the chaos
which give rise in this case. In Sec.3 a more wide class of potentials
less steep than quadratic is studied. In Sec.4 we discuss the
transition to regular dynamics in the presence of ordinary matter in
addition to the scalar field for potentials under consideration.

\section {Asymptotically flat potentials and the merging of bounce
intervals}
We will use the units in which $m_{P}/\sqrt{16\pi}=1$ for presenting
our numerical results, because in these units most of the interesting events
occur for the range of parameters of the order of unity.

 We start with the potential
\begin{equation}
V(\varphi)=M_{0}^{4}(1-exp(-\frac{\varphi^2}{\varphi_{0}^{2}})),
\end{equation}
where $M_0$ and $\varphi_0$ are parameters. $M_0$ determines the asymptotical
value of the potential for $\varphi \to \pm \infty$.

It can be easily checked from the equations of motion that multiplying
the potential to a constant (i.e. changing the $M_0$) leads only to
rescaling $a$. So, this procedure do not change the
chaotic properties of our dynamical system. On the contrary,
this system appear
to be very sensitive to the value of $\varphi_0$. We plotted in Fig.1.
the $\varphi=0$ cross-section of bounce intervals depending on $\varphi_0$.
This plot represents a situation, qualitatively different from studied
previously for potentials like $V \sim \varphi^2$ and steeper. Namely,
the bounce intervals can merge.

\begin{figure}
\epsfxsize=\hsize
\centerline{{\epsfbox{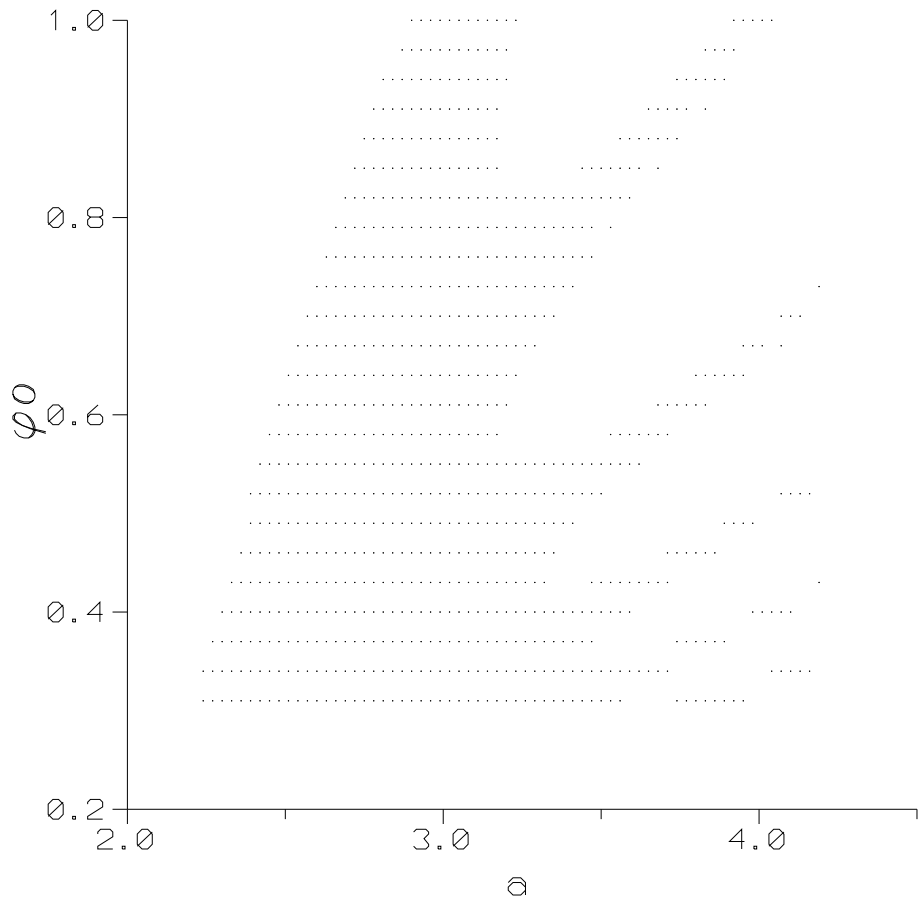}}}
\caption{
The $\varphi=0$ cross-section of the bounce intervals for the potential
(6) depending on $\varphi_0$. Consecutive merging of $5$ first intervals
can be seen in this range of $\varphi_0$.
}
\end{figure}

Let us see more precisely what does it means. For $\varphi_0>0.82$ the picture
is qualitatively as for a massive scalar field - trajectories from 1-st
interval have a bounce with no $\varphi$-turns before it,
trajectories which have initial point of maximal expansion between
1-st and 2-nd intervals fall into a singularity after one $\varphi$-turn,
 those from
2-nd interval have a bounce after 1 $\varphi$-turn and so on. For $\varphi_0$
a bit smaller than the first merging value the 2-nd interval contains
trajectories with 2 $\varphi$-turns before bounce, the space between
1-st interval (which is now the product of two merged intervals)
and the 2-nd one contains trajectories falling into a singularity
after two $\varphi$-turns. There are no trajectories going to a singularity
with exactly one $\varphi$-turn.
Trajectories from the 1-st
interval can experience now
a complicated chaotic behavior which can not
be described in as similar way as above.

With $\varphi_0$ decreasing further, the process of interval merging
being to continue leading to growing chaotisation of trajectories.
When $n$ intervals merged together, only trajectories with at
least $n$ oscillations of the scalar field before falling into
a singularity are possible. Those having exactly $n$ $\varphi$-turns
have their initial point of maximal expansion between 1-st bounce interval
and the 2-nd one (it now contains trajectories having a bounce after
$n$ $\varphi$-turns). For initial values of the scale factor larger then
those
from the 2-nd interval, the regular
quasiperiodic structure described above is restored.

 Numerical analysis shows also
that the fraction of very chaotic trajectories as a function of
$\varphi_0$ grows rapidly with $\varphi_0$ decreasing below the first
merging value. To illustrate this point we plotted in Fig.2 the number of
trajectories which do not fall into a singularity during first $50$
oscillations of the scalar field $\varphi$.  We do not include
trajectories with the next point of maximal expansion located outside
the 2-nd (or the 1-st one, if merging occurred) interval, so all
counted trajectories avoid a singularity during this sufficiently long
 time interval due to their extreme chaoticity, but not due to reaching
the slow-roll regime. The initial value of $a$ vary in the range of
the first two intervals before and after merging
with the step
$0.002$. Before merging, the measure of so chaotic trajectories is
extremely low and they are undistinguisheble on our grid. When $\varphi_0$
becomes slightly low than the value of the first merging, this number
begin to grow rather rapidly and for $\varphi_0 \sim 0.6$ near $10 \%$ of
trajectories from the 1-st interval experience at least 50 oscillation
before falling into a singularity.

\begin{figure}
\epsfxsize=\hsize
\centerline{{\epsfbox{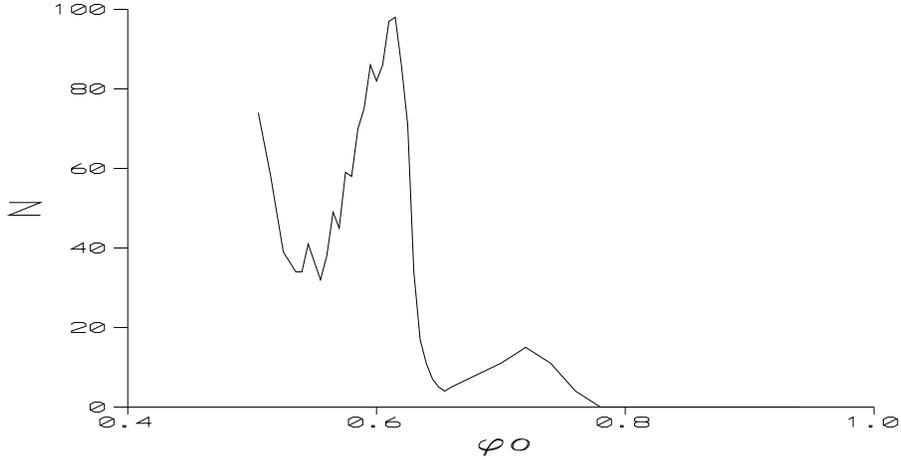}}}
\caption{
Number $N$ of trajectories do not falling into a singularity during
$50$ oscillating times for the potential (6) depending on the parameter
$\varphi_0$. The scale factor of the initial maximal expansion point varies
in the range of the 1-st and 2-nd intervals which merge at $\varphi_0=0.82$.
Total number of trajectories is equal to $1000$.
}
\end{figure}

We recall that for a simple massive scalar field potential only
$\sim 10^{-2}$ trajectories in the same range of the initial scale factors
have at least one bounce. Fraction
of trajectories not falling into a singularity after only one bounce
is about one hundred times less and so on. The common numerical
calculation accuracy is unsufficient for distinguishing even the sole
trajectory with 50 oscillations and $a$ being in the range of first two
intervals.

In contrast to this, the chaos
for the potential (6) is really significant. Detail of intervals merging
including the description out of $\varphi=0$ cross-section require further
analysis.

For large  initial $a$ the configuration of bounce intervals
for potential (6) looks like the configuration for
a massive scalar field potential with the effective
mass easily derived from (6): $m_{eff}=(\sqrt{2} M_{0}^{2})/\varphi_0$.
The periods of corresponding structures coincides with a good accuracy
though the widths of the intervals for the potential (6) is
bigger then for $V=(m_{eff}^2 \varphi^2)/2$.

\section{Damour-Mukhanov potentials}
The very chaotic regime described above is possible also for potentials,
which are not asymptotically flat, if the potential growth is slow enough.
We will illustrate this point describing
  a particular (but rather wide) family of
potentials having power-low behavior -- Damour - Mukhanov potentials
\cite{Damour}.
They was originally introduced to show a possibility to have an
inflation behaviour without slow-roll regime. After, various issues on
inflationary dynamics \cite{Liddle}
and growth of perturbation \cite{Taruya,Cardenas}
for this kind of scalar
field potential was studied.

The explicit form of Damour-Mukhanov potential is
\begin{equation}
V(\varphi)=\frac{M_{0}^{4}}{q} \left[ \left(1+\frac{\varphi^2}
{\varphi_{0}^{2}} \right)^{q/2}-1 \right],
\end{equation}
with three parameters --$M_0$, $q$ and $\varphi_0$.

For $\varphi \ll \varphi_0$ the potential looks like the massive one with the
effective mass $m_{eff}=M_{0}^{2}/\varphi_0$. In the opposite case of large
$\varphi$ it grows like $\varphi^q$.

As in the previous section, the chaotic behavior does not depend on
$M_0$.  So, we have a two-parameter family of potentials with different
chaotic properties. Numerical studies with respect to possibility of
bounce intervals merging shows the following picture (see Fig.3): for
a rather wide range of $q$ there exists a corresponding critical value
of $\varphi_0$ such that for $\varphi_0$ less than critical, the very
chaotic regime exists. Increasing $q$ corresponds to decreasing
critical $\varphi_0$.

\begin{figure}
\epsfxsize=\hsize
\centerline{{\epsfbox{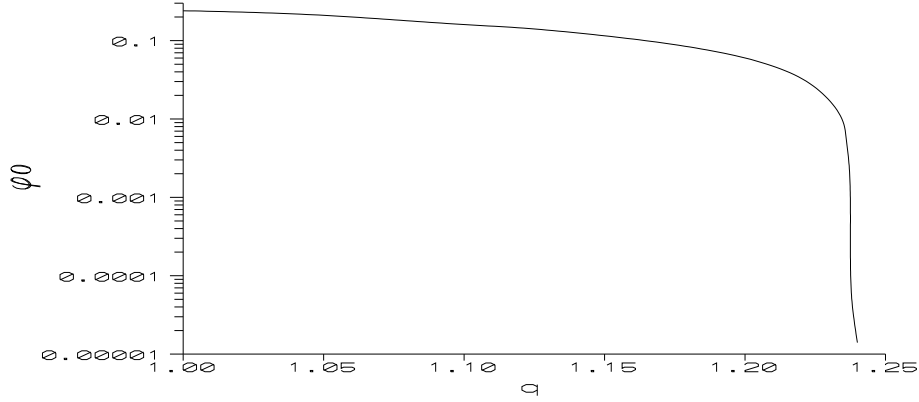}}}
\caption{
The value $\varphi_0$ of the potential (7) corresponding to the
first merging of the bounce intervals depending on $q$.
}
\end{figure}

 Surely, since this regime is absent for quadratic and more
steep potentials, $q$ must at least be less than $2$. We can see clearly
the very chaotic regime for $q< 1.24$.
 The case $q=1.24$ lead to strong chaos for $\varphi_0<1.4 \times 10^{-5}$ and
the critical $\varphi_0$ decreases with increasing $q$ very sharply at this
point. We did not investigated further these extremely small values of
$\varphi_0$, because the physical meaning of such kind of potential is
very doubtful.

\section{The influence of a hydrodynamical matter}
In this section we add the perfect fluid with the equation of state
$P=\gamma \epsilon$.
The equation of motion are now
\begin{equation}
\frac{m_{P}^{2}}{16 \pi}\left(\ddot{a} + \frac{\dot{a}^{2}}{2 a}
    + \frac{1}{2 a} \right) +\frac{a \dot{\varphi}^{2}}{8}
-\frac{V(\varphi)}{4}
            - \frac{Q}{12 a^{p+1}}(1-p) = 0
\end{equation}
\begin{equation}
\ddot{\varphi}
    + \frac{3 \dot{\varphi} \dot{a}}{a}+ V'(\varphi) = 0.
\end{equation}
with the constraint
\begin{equation}
-\frac{3}{8 \pi} m_{P}^2 (\dot{a}^{2} + 1)
+\frac{a^{2}}{2}\left(\dot{\varphi}^{2} + 2 V(\varphi)\right) +
                \frac{Q}{a^{p}} = 0.
\end{equation}
Here $p=1+3 \gamma$, $Q$ is a constant from the equation of motion for matter
which can be integrated in the form
\begin{equation}
    E a^{p+2} = Q = const.
\end{equation}

In Ref.\cite{ournew} it was shown that addition of a hydrodynamical matter
to the scalar field with a potential $V(\varphi)=m^2 \varphi^2/2$ can kill
the chaos. Here we extend this analysis to less steep potentials.
Some of our results are illustrated in Fig.4. It is interesting that
increasing $Q$ acts as increasing $\varphi_0$. In Fig.4 three intervals
merged at $Q=0$.  When $Q$ increases, the 3-d and 2-nd interval
consecutively separate and we return to the chaos typical for
$V(\varphi)=m^2 \varphi^2/2$. With the further increasing of $Q$ the chaos
disappear in a way discussed in \cite{ournew}.

\begin{figure}
\epsfxsize=\hsize
\centerline{{\epsfbox{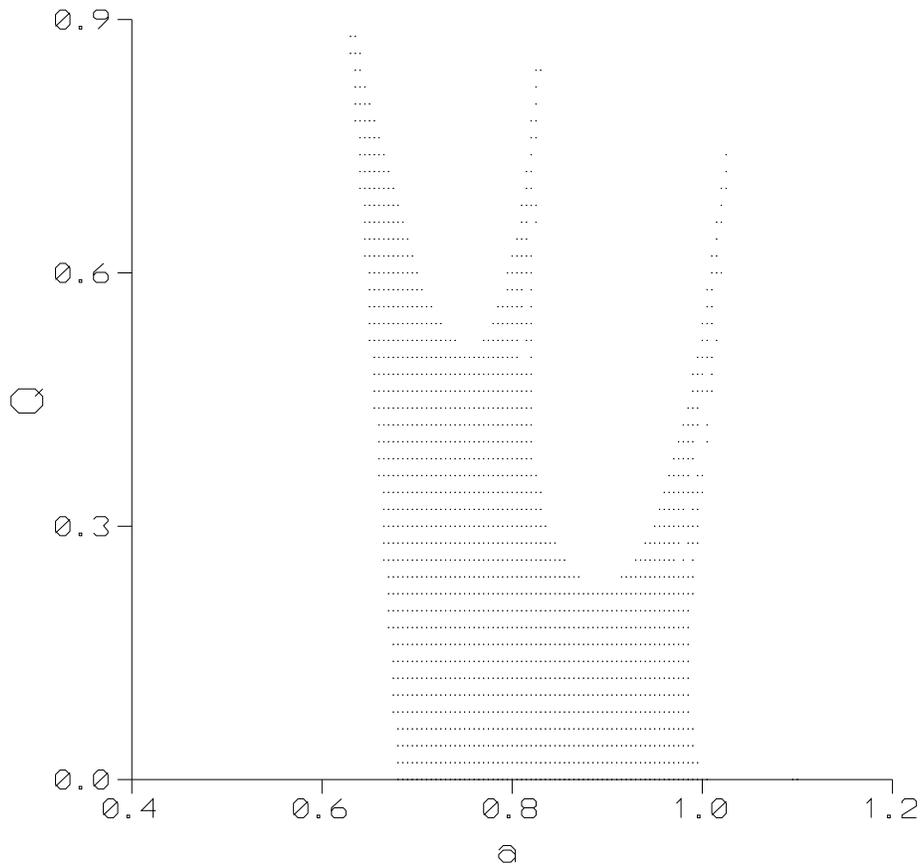}}}
\caption{
The $\varphi=0$ cross-section of the bounce intervals for the Damour-
Mukhanov potential with $q=1.0$, $\varphi_0=0.1$ depending of the $Q$. For
$Q=0$ three intervals merges. This plot shows the consecutive separation
and further disappearance of the bounce intervals with $Q$ increasing.
}
\end{figure}

The value of $Q$ corresponding to the chaos disappearing is in general
bigger for the less steep potentials with the same effective mass
. In Fig.5.(a) this values  are plotted
for Damour-Mukhanov potentials and the perfect fluid with $\gamma=0$
($p=1$).
 This particular $\gamma$ is chosen for a mathematical reason. Namely,
it can be seen from (8)-(10) that in this case only the constraint
equation is changed in comparison with the initial system (1)-(3). In
other words, dynamical equations describing FRW universe with a scalar
field and dust matter are formally equivalent to those for scalar field
only but with nonzero value of the conserved energy. So, our figure
describes not only the physical system under consideration, but also
general mathematical properties of (1)-(2).

 We recall that for the case $V(\varphi)=m^2 \varphi^2/2$ (which is
 equivalent to the Damour-Mukhanov potential with $q=2$, the
corresponding mass is equal to the effective one
$m_{eff}=M_{0}^{2}/\varphi_0$) the chaos disappear for $Q m > 0.023
m_{P}$ \cite{ournew}. To compare with this value, we plotted in
Fig.5(a) the values of $Q m_{eff}$ leading to ceasing of the chaos with
respect to $\varphi_0$ for several $q$.  In units we used the case
$q=2$ corresponds to horizontal line $Q m_{eff} = 1.15$. All other
curves have this value as an asymptotic one for large $\varphi_0$. With
decreasing $\varphi_0$ the value $Q m_{eff}$ increases with the rate
which is bigger for less steep potentials.

\begin{figure}
\epsfxsize=\hsize
\centerline{{\epsfbox{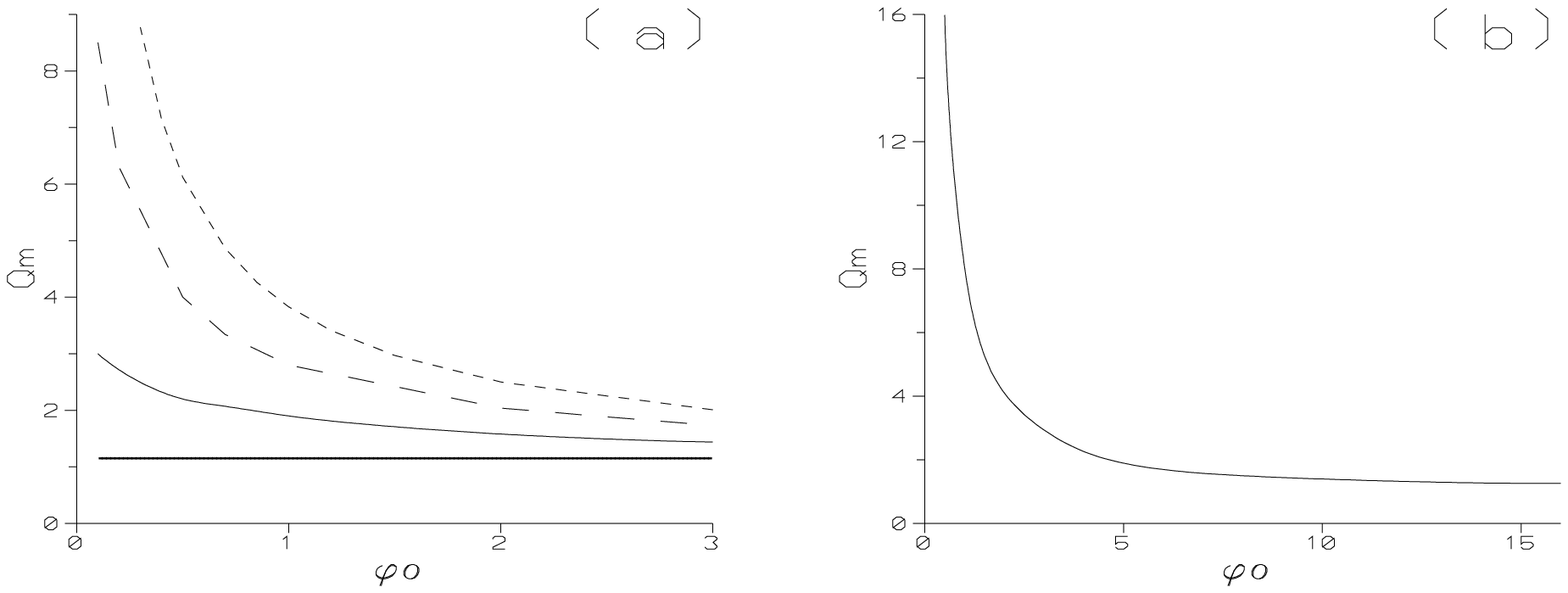}}}
\caption{In Fig.5(a)
the values of $Q m_{eff}$ killing the chaos for potentials
(7) depending on $\varphi_0$ are plotted
 for several $q$: $q=2$ (bold line), $q=1.5$
(solid curve), $q=1.0$ (long-dashed curve), $q=0.5$ (short-dashed curve).
In Fig.5(b)
the values of $Q m_{eff}$ killing the chaos for potentials
(6) depending on $\varphi_0$ are plotted.
}
\end{figure}

In Fig.5(b) the analogous curve is plotted for asymptotically flat
potential (6). The value $Q m_{eff}$ for large $\varphi_0$ is the same.
For small $\varphi_0$ we can estimate $Q$ killing the chaos as the
value corresponding to disappearance of the Euclidean region for a flat
potential $V(\varphi)=M_{0}^{4}$. It can be easily obtained from (4) that
in this case the Euclidean region disappears at $$ Q=\frac{1}{8 \sqrt{2
\pi^3}} \frac{m_{P}^{3}}{M_{0}^{2}} $$ and for bigger $Q$ any bounce
become impossible. As the potential (6) differs significantly from the
flat one only for $\varphi$ less then $\varphi_0$, this approximation appear to be
good enough for small $\varphi_0$. In the our units it correspond to the
curve $Q m_{eff}=8/\varphi_0$. Intermediate values of $\varphi_0$ represent
smooth transition between these two asymptotic behaviors.

\section*{Acknowledgments}

This work was supported by
Russian Basic Research Foundation via grant
No 99-02-16224.

\end{document}